\documentclass[prb,showpacs,twocolumn,aps]{revtex4}
\usepackage{graphicx}
\newcommand{\be}{\begin{equation}}
\newcommand{\ee}{\end{equation}}
\newcommand{\bea}{\begin{eqnarray}}
\newcommand{\eea}{\end{eqnarray}}
\newcommand{\bwt}{\begin{widetext}}
\newcommand{\ewt}{\end{widetext}}
\begin{document}
\title{Transport in vortex state of $d$-wave superconductors
at zero temperature: Wiedemann-Franz violation
}
\author{Wonkee Kim,$^{1}$ F. Marsiglio,$^{1}$ and  J. P. Carbotte$^{2}$}
\affiliation{
$^{1}$Department of Physics, University of Alberta, Edmonton, Alberta,
Canada, T6G~2J1\\
$^{2}$Department of Physics and Astronomy,
McMaster University, Hamilton,
Ontario, Canada, L8S~4M1}
\begin{abstract}
We show that the Wiedemann-Franz law is violated at zero temperature in the vortex
state of a $d$-wave BCS superconductor with isotropic impurity scattering.
We use a semiclassical approach to include the Doppler shift experienced by
the quasiparticles due to the circulating supercurrents and consider as well
the Andreev scattering from an array of vortices assumed to be randomly distributed.
We also show that the vertex corrections to the electric conductivity, which can be
large when there is significant anisotropy in the impurity scattering,
become unimportant as the magnetic field is increased.
For the thermal conductivity, the corrections remain negligible as in the
absence of a magnetic field.
\end{abstract}
\date{\today}
\maketitle

\section{introduction}

The Wiedemann-Franz (WF) law is a hallmark of metallic behavior. This law
states that the ratio of the thermal conductivity ($\kappa$) to the dc optical 
conductivity ($\sigma$), is given by a universal number times 
the temperature $(T)$.
The Lorenz number is defined by $L \equiv \kappa/(T\sigma)$, and is expected to
equal $\pi^2/(3e^2)$ (we use $\hbar \equiv k_B \equiv 1$). This law is well obeyed
in most metals, and can be theoretically justified well beyond the case of a
simple metal.\cite{ashcroft}  

Recently several experiments have been designed to test the validity of the
WF law, particularly at very low temperature, in the
cuprate superconductors. The law is well obeyed in overdoped 
Tl$_{2}$Ba$_{2}$CuO$_{6+\delta}$ which is
forced into its normal state through application of a magnetic field.\cite{proust}
A similar result was also found in optimally doped 
Bi$_{2+x}$Sr$_{2-x}$CuO$_{6+\delta}$.\cite{bel}
These results clearly favor the existence of ordinary quasiparticles
in these materials and do not support the spin-charge separation
picture of the Luttinger liquid.\cite{kivelson,anderson,senthil} 
In this picture collective modes
replace the quasiparticles with spinons carrying the spins and holons
the charges, and the WF law does not hold.\cite{kane}
On the other hand the normal state of a slightly underdoped sample of 
Pr$_{2-x}$Ce$_{x}$CuO$_{4-y}$ 
studied by Hill {\it et al.}\cite{hill} revealed serious violations of
the WF law, which indicates the existence of an exotic normal state.
The data puts severe constraints on the state involved. For example,
formation of a competing hidden ordered state such as the $d$-density wave
(DDW) state is not expected to lead to a violation of the WF law.\cite{kim}

In this paper we will confine our discussion to the
superconducting state with a $d$-wave gap symmetry.
In particular we will study the mixed state where vortices are created by
the application of an external magnetic field $H$ perpendicular to the two
dimensional CuO$_{2}$ plane.

Near the critical temperature $T_{c}$, the cuprates exhibit strong inelastic 
scattering. The inelastic scattering rate is of order a few times $T_{c}$.\cite{tanner}
The large peak seen in the microwave conductivity\cite{bonn,hosseini}
as a function of temperature is widely interpreted as evidence for the
collapse of the inelastic scattering\cite{nuss,schachinger} as
$T$ is reduced below $T_{c}$. 
The idea is that the development of a superconducting gap suppresses the
low frequency density of bosons which causes the superconductivity. This
suppression is a general characteristic of any electronic mechanism,
in which the boson is a collective mode of the superconducting electrons.
Consequently for
$T\ll T_{c}$ elastic impurity scattering will dominate even in the purest samples in which
the quasiparticle mean free path can be of order one micron.\cite{hosseini}
Therefore in the low temperature regime, which is the region of interest here, it is
sufficient, as a first approximation, to ignore the specifics of the mechanism
involved in the inelastic scattering.
A $d$-wave BCS formulation which includes impurity scattering will suffice.

For a $d$-wave gap symmetry both the quasiparticle density of states and the
effective impurity scattering rate acquire an 
important frequency $(\omega)$ dependence.
These essential $\omega$ dependences can lead to new phenomena. For example
impurities modify, in an essential way, the quasiparticle density of states
$N(\omega)$ in the limit $\omega\rightarrow0$.\cite{hirschfeld1,hirschfeld2} 
In the pure case the density of states approaches zero linearly; however, with
impurity scattering it takes on a finite value at zero frequency which
leads directly to the concept of the universal value
of the conductivity at $\omega=0$ and $T=0$, 
independent of impurity content.\cite{lee} This behavior also holds in the case of
the thermal conductivity and has been verified in many experiments.
\cite{taillefer,chiao,nakamae,takeya,sutherland} However, 
the WF law still holds in this limit.

If finite temperature is considered, the WF law is violated.\cite{graf}
Violations can be either positive or negative (the Lorenz number is greater 
or less than one), dependent on the details of the impurity potential. 
In the unitary limit, the Lorenz number can
increase rapidly with increasing $T$ while in the Born limit (weak scattering)
it can decrease. In pure samples it drops to one half its original $(T=0)$ value
within a few degrees Kelvin.\cite{graf} 
In the simplest approximation, $\kappa$
and $\sigma$ both
depend on the overlap of thermal factors with a product of the density of
quasiparticle states $N(\omega)$ and the scattering time $\tau(\omega)$. The thermal
factors for the electrical case are peaked about $\omega=0$ while for the thermal
case they peak at higher frequencies
of order $T$. This means that a different frequency range
of $N(\omega)\tau(\omega)$
is most importantly sampled in the two cases and the WF law no longer holds
at finite temperature.\cite{wu}
The Lorenz number
can be smaller or greater than its conventional
value depending on the behavior of
$N(\omega)\tau(\omega)$ as a function of $\omega$.
For the superconducting state this important function $N(\omega)\tau(\omega)$
is different for the electrical and thermal conductivity because of differences
in coherence factors but the physics described above for the normal state continues
to hold.

Recent extensions of the theory at $T=0$ to include vertex corrections, which is
necessary when the impurity potential is anisotropic, have led to modifications of
the above picture. Durst and Lee (DL) found that in these circumstances there
can be important corrections to the electrical conductivity while
the thermal conductivity is almost unaltered.\cite{durst} 
Consequently the WF law no longer holds
even at $T=0$ and the Lorenz number is found to be reduced. 
The amount of reduction
depends on the impurity potential $V_{i}$ (Ref.\cite{schachinger2})
and is largest for the unitary limit
$(V_{i}\rightarrow\infty)$.
The physics underlying this phenomenon
is well described in Ref.\cite{durst} and has its origin in the fact that the
thermal conductivity depends on the group velocity of the quasiparticles while its
electrical counterpart is sensitive to their Fermi velocity.

In this paper we want to consider the effect that a magnetic field $(H)$ applied 
perpendicular to the two dimensional $d$-wave superconducting plane will have on
the WF law and on the vertex corrections for the in-plane transport. It is well
known\cite{volovik,kubert1,kubert2,vekhter} 
that the application of $H$ will increase the density of states 
at zero frequency $N(0)$ and thus increase both
$\sigma$ and $\kappa$. When $H$ is larger than the lower critical field $H_{c1}$ 
but still much less than the upper critical field $H_{c2}$, vortices form
and these will provide 
supercurrents as well as
an added scattering mechanism for the quasiparticles.
Here we treat effects of the supercurrents semiclassically and 
this additional scattering in a phenomenological model as Andreev
scattering\cite{cleary,yu,franz} which can be different for charge and heat
transport. 
We found 
the WF law is violated at $T=0$
in the mixed state of a BCS $d$-wave superconductor. This result holds even when
the vertex corrections for anisotropy are not taken into account, {\it i.e.}
at the level of the bare bubble diagrams in the treatment of the correlation
functions involved.
We have also studied the influence of an external magnetic field on the 
vertex corrections.  For the case of the electrical conductivity, it is 
found that the vertex corrections present at zero $H$ can be rapidly 
suppressed as $H$ increases. For the thermal conductivity,
the vertex corrections are never important.

This paper is organized as follows: In Section II, we describe the formalism
for $d$-wave superconductors with impurities in the vortex state. As a
first approximation, Matthiessen's rule is applied to consider the impurity
and vortex scattering. We also study self-consistently, in Section III,
effects of the magnetic field on the impurity scattering, which is beyond
Matthiessen's rule. In Section IV,
magnetic field effects on the vertex corrections
to $\sigma$ and $\kappa$ is presented. Section V is devoted
to conclusions. 

\section{theoretical approach}

The calculation of the thermal and dc conductivity proceeds in the standard way
through the evaluation of the appropriate current-current correlation
functions for heat and charge current, respectively. Formally, the required
correlation function is written in Nambu notation ($2\times2$ matrix notation)
and an analytic continuation from the Matsubara to real frequencies is needed
before the zero frequency limit is taken. 
The superconducting matrix Green's function
in the Nambu notation takes the form
${\hat G}({\bf k},i\omega_{n})=(\tilde{\omega}{\hat\tau}_{0}
+\Delta_{\bf k}{\hat\tau}_{1}+\xi_{\bf k}{\hat\tau}_{3})/
(\tilde{\omega}^{2}-\xi^{2}_{\bf k}
-\Delta^{2}_{\bf k}),$
where ${\hat\tau}$'s are the Pauli matrices in spin space,
$\Delta_{\bf k}$ a $d$-wave order parameter, and $\xi_{\bf k}$ is
the electronic
energy dispersion in the normal state.
The correlation function can be 
expanded into a product of two Green's functions and a vertex function. 
At the lowest level of approximation, the vertex is neglected 
and only the bare bubble is retained. While we start with this approximation
later we will generalize our work to include the vertex corrections as in
Ref.\cite{durst}; these vanish when the impurity scattering is isotropic, but are
required when the scattering is anisotropic.

To include the effects of an external magnetic field $H$, for
$H_{c1}\lesssim H \ll H_{c2}$, we employ the semiclassical approximation 
which includes, in the single particle Green's functions, 
the Doppler shift\cite{volovik,kubert1,kubert2,vekhter,hussey} 
associated with the circulating supercurrent
around a vortex core. This is accomplished formally by changing the Matsubara
frequencies $i\omega_{n}$ in the single particle Green's function to
$i\omega_{n}-{\bf v}_{s}\cdot{\bf k}$, where ${\bf v}_{s}$ is the velocity field
of the supercurrent and ${\bf k}$ is the momentum of a quasiparticle on the
Fermi surface. Now the new Green's function depending on ${\bf v}_{s}\cdot{\bf k}$
is inhomogeneous in space variable ${\bf r}$ through the dependence of
the velocity field ${\bf v}_{s}({\bf r})$ on position ${\bf r}$ measured from
the core of a vortex. Thus thermal and transport coefficients become
local quantities which need to be averaged over a vortex unit cell,
assumed to be a circle of radius $R$ for simplicity.
The radius $R$ is related to the coherence length $\xi_{0}$ and $H_{c2}$
through $2R=\xi_{0}\sqrt{2\pi}a^{-1}\left(H_{c2}/H\right)^{1/2}$, where
$a$ is a geometrical factor of order unity.
The average of any quantity ${\cal F}({\bf r})$ is then
\be
{\cal F}(H)=\frac{1}{\pi R^{2}}\int{}d{\bf r}\;{\cal F}({\bf r})\;.
\ee
We will assume that $\xi_{0}\ll R$ and that the circulating current velocity
field can be taken to drop as $1/r$ throughout the vortex unit cell. For 
a magnetic field applied perpendicular to the CuO$_{2}$
plane, the Doppler shift ${\bf v}_{s}\cdot{\bf k}$ takes the form
$E_{H}/\rho$ with $\rho=r/R$ and $E_{H}=av_{f}/2\sqrt{\pi H/\Phi_{0}}$, where
$v_{f}$ is the Fermi velocity and
$\Phi_{0}$ is the flux quantum related to $H_{c2}$ by 
$\Phi_{0}=2\pi\xi^{2}_{0}H_{c2}$.

Impurities can be introduced into the calculations through the impurity self
energy term which is to be evaluated self-consistently in a t-matrix approximation.
The retarded self energy $\Sigma_{i,ret}$ is given by
\be
\Sigma_{i,ret}({\tilde\omega})=\frac
{\Gamma G_{0}({\tilde\omega})}{c^{2}-G^{2}_{0}({\tilde\omega})}\;,
\label{Sigma}
\ee
where $G_{0}(\omega)=[2\pi N(0)]^{-1}
\sum_{\bf k}\mbox{Tr}[{\hat G}_{ret}({\bf k},\omega)]$ 
with 
${\hat G}_{ret}$ the retarded Green's function. The summation over ${\bf k}$
is to be carried out over the two dimensional Brillouin zone.
In Eq.~\ref{Sigma}, $\Gamma$ is related to the impurity density $n_{i}$ by
$\Gamma=n_{i}/\left[\pi N(0)\right]$ and $c$ is the strength of the impurity
potential $V_{i}$; specifically $c=1/(\pi N(0)V_{i})$. The unitary limit
corresponds to $V_{i}\rightarrow\infty$, i.e. $c\rightarrow0$, while the
Born limit corresponds to $V_{i}\rightarrow0$ $(c\rightarrow\infty)$. As we have done 
in our previous work on the thermal conductivity,\cite{kim2} in the mixed state
we assumed as a first approximation that $\Sigma_{i,ret}$ remains unmodified
by an external magnetic field. That is, appealing to Matthiessen's rule, we 
add on to the term $\mbox{Im}\Sigma_{i,ret}\equiv\gamma(\omega)$ a term which
describes phenomenologically the Andreev scattering of the quasiparticles
off the vortex core. A simple expression for the quasiparticle-vortex scattering
has been given by Yu {\it et al.}\cite{yu} Examination of their formula
for the geometry used here ($H\perp$CuO$_{2}$ plane) 
reveals that it is proportional
to the magnetic energy $E_{H}$ as also argued through dimensional analysis
in Ref.\cite{franz} The constant of proportionality $b$ is to be determined 
by comparison with experimental data as done in our previous work.\cite{kim2}
This linear dependence of the vortex scattering on $E_{H}$ implies
that the vortex cores behave like scattering centers. This means
that we consider two effects due to a vortex; 
the Doppler shift due to the supercurrent as well as the vortex scattering
due to the core.
The total scattering rate denoted by $\gamma_{tot}(\omega)$ is then given by
\be
\gamma_{tot}(\omega)=\gamma(\omega)+bE_{H}\;,
\label{g_tot}
\ee
where the first term $\gamma(\omega)$ is due to impurities and
the second term $bE_{H}$ is due to vortex scattering. Later we will
include the effect of $H$ on the impurity scattering itself.

A simplification in the required algebra results when the limit of low $T$
is considered. For $T\ll\Delta_{0}$ (amplitude of a $d$-wave gap), 
it is appropriate
to make a nodal approximation \cite{durst} 
for the summation over the Brillouin zone. 
Application of this approximation to $G_{0}(\omega)$ in Eq.~(\ref{Sigma}) leads
to
\be
G_{0}(\tilde{\omega})\simeq
\frac{2}{\pi}\left[
\frac{\tilde{\omega}}{\Delta_{0}}
\ln\frac{\tilde{\omega}}{4\Delta_{0}}-
i\frac{\pi}{2}\frac{\tilde{\omega}}{\Delta_{0}}\right]\;,
\label{G_0}
\ee 
where a cut-off of $4\Delta_{0}$ has been applied to the real part
of $G_0$ in Eq.~(\ref{G_0}). In a similar way, application of the nodal
approximation to the expression for the thermal conductivity
$\kappa(T,H)$ leads to
\bwt
\be
\frac{\kappa (T,H)}{T}=\frac{1}{\pi^{2}}\frac{1}{T^{2}}
\left(\frac{v_{f}}{v_{g}}+\frac{v_{g}}{v_{f}}\right)
\int{}d\epsilon {\cal P}(\epsilon)\int{}d\omega\;
\omega^{2}\left(-\frac{\partial f}{\partial\omega}\right)
{\cal A}_{\kappa}(\omega,\epsilon)\;,
\label{kappa_TH}
\ee 
\ewt
where $v_{f(g)}$ is the Fermi (gap) velocity and 
$f(\omega)=\left[1+\exp(\omega/T)\right]^{-1}$ is the Fermi-Dirac
thermal factor. 
The thermal conductivity in the plane $\kappa_{xx}=\kappa/2$ in the nodal
approximation. This relation also holds for the electrical conductivity.
In deriving Eq.~(\ref{kappa_TH}) we have introduced a vortex distribution function 
${\cal P}(\epsilon)$ through the equation
\be
{\cal P}(\epsilon)=\frac{1}{\pi R^{2}}\int{}d{\bf r}\;\delta\left(
\epsilon-{\bf v}_{s}({\bf r})\cdot{\bf k}\right)\;,
\ee 
where the integration is over the vortex unit cell. The actual thermal conductivity
is obtained by averaging the local quantity over the unit cell.
In Eq.~(\ref{kappa_TH}), the function ${\cal A}_{\kappa}(\omega,\epsilon)$ can be 
written as
\be
{\cal A}_{\kappa}(\omega,\epsilon)=
1+\left(\frac{\bar{\omega}-\epsilon}{\gamma_{t}(\omega)}+\frac{\gamma_{t}(\omega)}
{\bar{\omega}-\epsilon}\right)
\arctan\left(\frac{\bar{\omega}-\epsilon}{\gamma_{t}(\omega)}\right)\;,
\label{Ak}
\ee
where $\bar{\omega}=\omega-\mbox{Re}\left[\Sigma_{ret}\right]$. Several choices
can be made for the vortex distribution function and results depend somewhat
on the choice. However, qualitative results are not sensitive to the choice.
To be specific we report here the results based on the Gaussian distribution
\be
{\cal P}(\epsilon)=\frac{1}{\sqrt{\pi}E_{H}}\exp\left(-\frac{\epsilon^{2}}
{E^{2}_{H}}\right)\;.
\ee
Note that for $H=0$ we can take $\epsilon=0$ in Eq.~(\ref{Ak}), namely,
${\cal A}_{\kappa}(\omega,0)$. If we further take $T\rightarrow0$, we obtain
\be
\frac{\kappa_{00}}{T}=\frac{2}{3}\left(\frac{v_{f}}{v_{g}}+\frac{v_{g}}{v_{f}}\right)\;.
\label{k00}
\ee
This value is the well-known expression for the universal limit of the thermal
conductivity independent of the impurity scattering rate 
at zero frequency $(\gamma_{00})$.
If instead we had taken $T\rightarrow0$ first in Eq.~(\ref{kappa_TH}),
we would have 
\be
\lim_{T\rightarrow0}\frac{\kappa (T,H)}{T}=\frac{1}{3}
\left(\frac{v_{f}}{v_{g}}+\frac{v_{g}}{v_{f}}\right)
\int{}d\epsilon {\cal P}(\epsilon)
{\cal A}_{\kappa}(0,\epsilon)\;,
\label{kappa_T0}
\ee
where
\be
{\cal A}_{\kappa}(0,\epsilon)=
1+\left(\frac{\epsilon}{\gamma_{t}(0)}
+\frac{\gamma_{t}(0)}{\epsilon}\right)
\arctan\left(\frac{\epsilon}
{\gamma_{t}(0)}\right)\;.
\ee
Note that in this case only the zero frequency limit of $\gamma_{t}$ enters. At any
finite $T$, however, the $\omega$ dependence of $\gamma_{t}(\omega)$ becomes very
important, and the $T$ dominated region $T\gg\gamma_{t}(0)$ can be quite different
from the impurity dominated region for which $T\ll\gamma_{t}(0)$. Finally
we note that taking $H \rightarrow 0$ in Eq.~(\ref{kappa_T0}) again gives the value of 
the universal limit as it should.

Similar algebra gives the expression for the zero frequency value of the optical 
conductivity $\sigma(T,H)$, which takes the form
\bwt
\be
\sigma(T,H)=\frac{e^{2}}{\pi^{2}}\left(\frac{v_{f}}{v_{g}}\right)
\int{}d\epsilon {\cal P}(\epsilon)\int{}d\omega
\left(-\frac{\partial f}{\partial\omega}\right)
{\cal A}_{\sigma}(\omega,\epsilon)\;,
\label{sigma_TH}
\ee
\ewt
where
\be
{\cal A}_{\sigma}(\omega,\epsilon)=2\left[1+\frac{\bar{\omega}-\epsilon}{\gamma_{t}(\omega)}
\arctan\left(\frac{\bar{\omega}-\epsilon}{\gamma_{t}(\omega)}\right)\right]\;.
\ee
In Eq.~(\ref{sigma_TH}), $e$ is the electron charge. 
We note that the difference between ${\cal A}_{\kappa}$ and ${\cal A}_{\sigma}$
is traced to the different coherence factors that enter, respectively, the
electrical and thermal conductivity. In terms of the spectral functions
$A({\bf k},\omega)$ and $B({\bf k},\omega)$ associated, respectively, with
the diagonal and off-diagonal part of Green's function, the optical (thermal) case
involves $A^{2}({\bf k},\omega) \pm B^{2}({\bf k},\omega)$, respectively.\cite{kim}
As $T\rightarrow0$, we obtain
\be
\sigma(0,H)=
\frac{e^{2}}{\pi^{2}}\left(\frac{v_{f}}{v_{g}}\right)
\int{}d\epsilon {\cal P}(\epsilon)
{\cal A}_{\sigma}(0,\epsilon)\;.
\label{sigma_0H}
\ee
If we also take $H=0$, we recover the well-known result of the universal value
of the dc conductivity
\be
\sigma_{00}=2\frac{e^{2}}{\pi^{2}}\left(\frac{v_{f}}{v_{g}}\right)\;.
\label{s00}
\ee 
Now for $H=0$ and $T\rightarrow0$, the 
Lorenz number $L=\kappa/T\sigma$ can be evaluated to test the WF law. Using
Eqs.~(\ref{k00}) and (\ref{s00}), we obtain
$L_{00}=\kappa_{00}/\left(T\sigma_{00}\right)\simeq\pi^{2}/\left(3e^{2}\right)$
assuming $v_{g}\ll v_{f}$. This is the universal constant normally associated
with the WF law. 

Remaining at $T=0$ we can consider the effects of 
a magnetic field $H$ on the WF law at the level of bare bubble {\it i.e.}
ignoring the vertex corrections.
The normalized thermal $(\kappa/\kappa_{00})$ and dc $(\sigma/\sigma_{00})$
conductivity can be written as
\be
\frac{\kappa}{\kappa_{00}}=
\frac{1}{\sqrt{\pi}}\int^{\infty}_{0}dx~e^{-x^{2}}\left[1+
\left(\frac{x}{{\Gamma}_{t}}+\frac{{\Gamma}_{t}}{x}\right)
\arctan\left(\frac{x}{{\Gamma}_{t}}\right)\right]
\label{norm_k}
\ee
and
\be
\frac{\sigma}{\sigma_{00}}=
\frac{2}{\sqrt{\pi}}\int^{\infty}_{0}dx~e^{-x^{2}}\left[1+
\frac{x}{{\Gamma}_{t}}
\arctan\left(\frac{x}{{\Gamma}_{t}}\right)\right]\;,
\label{norm_s}
\ee
where ${\Gamma}_{t}=\gamma_{t}(0)/E_{H}=
\gamma_{00}/E_{H}+b$ with $b=b_{\kappa}$ for the thermal conductivity
and $b=b_{\sigma}$ for the dc conductivity. These two quantities are not the same
for Andreev scattering off a vortex core. On physical grounds we expect
$b_{\kappa}>b_{\sigma}$ because Andreev scattering is more effective
for heat transport than for charge transport. Evaluation of these parameters
requires a microscopic mechanism for the quasiparticle-vortex scattering, which is
beyond the scope of this work. Here we effectively treat them as phenomenological 
parameters and present results for various values of $b$. The value of
$b_{\kappa}$ has been determined for two samples in our previous work\cite{kim2}
on the thermal conductivity. In the ultra pure sample of Hill {\it et al.}\cite{hill}
its value is about $b_{\kappa}=0.3$ while in the impure sample discussed by these same authors
we found that Andreev scattering is not important compared with the impurity scattering
so that the value of $b_{\kappa}$ is effectively zero.

A first result can be obtained analytically from Eqs.~(\ref{norm_k}) and (\ref{norm_s})
assuming no vortex scattering $(b_{\kappa}=b_{\sigma}=0)$. In the high field limit
$E_{H}\gg\gamma_{00}$ or ${\Gamma}_{t}\ll1$ we can expand 
$\kappa/\kappa_{00}$ and $\sigma/\sigma_{00}$ in terms of
$\gamma_{00}/E_{H}$ and obtain
\be
\frac{\kappa}{\kappa_{00}}=
\frac{\sqrt{\pi}}{4{\Gamma}_{t}}
-\frac{\sqrt{\pi}}{4}\left(2\ln{\Gamma}_{t}+\gamma_{e}-1\right)
\Gamma_{t}
\ee
where $\gamma_{e}\simeq0.577$ is the Euler number,
and
\be
\frac{\sigma}{\sigma_{00}}=\frac{\sqrt{\pi}}{2{\Gamma}_{t}}
+\frac{\sqrt{\pi}}{4}{\Gamma}_{t}\;.
\ee
Note that $E_{H}\sim\sqrt{H}$ indicating that
$\kappa/\kappa_{00}\sim
\sqrt{H}+{\cal O}\left(\ln(H)/\sqrt{H}\right)$
while $\sigma/\sigma_{00}\sim\sqrt{H}+{\cal O}\left(1/\sqrt{H}\right)$.
The Lorenz number is given by ($\Gamma_t << 1$): 
\be
L=\frac{\kappa}{T\sigma}=\frac{1}{2}L_{00}
\label{half_L}
\ee
where $L_{00}$ is the universal value of the Lorenz number. This means that
the Lorenz number at a high field is half its universal value, which is 
{\it a serious
violation of the WF law at zero temperature}. 
We should emphasize that to our knowledge this
law has always been obeyed in the past at $T=0$,
even in an exotic state such as the DDW state.\cite{kim,sharapov}
Our result represents a significant exception.

For finite but small vortex scattering ($b_{\kappa}\ll1$ and $b_{\sigma}\ll1$),
${\kappa}/{\kappa_{00}}$ and ${\sigma}/{\sigma_{00}}$ both saturate to values 
determined only by $b_{\kappa}$ and $b_{\sigma}$, respectively, in the limit $E_H >>
\gamma_{00}$,
\be
\frac{\kappa}{\kappa_{00}}=
\frac{\sqrt{\pi}}{4b_{\kappa}}
-\frac{\sqrt{\pi}}{4}\left[2\ln(b_{\kappa})+\gamma_{e}-1\right]
b_{\kappa}
\ee
and
\be
\frac{\sigma}{\sigma_{00}}=\frac{\sqrt{\pi}}{2b_{\sigma}}
+\frac{\sqrt{\pi}}{4}b_{\sigma}\;.
\ee
As $b_{\kappa}$ and $b_{\sigma}\rightarrow0$, we obtain
\be
\frac{L}{L_{00}}=\frac{b_{\sigma}}{2b_{\kappa}}\;.
\label{L_b}
\ee
Since we expect $b_{\sigma}$ to be less than $b_{\kappa}$, the Lorenz number
should be smaller than its universal value in this regime and Eq.~(\ref{L_b})
could be used to find the ratio $b_{\sigma}/b_{\kappa}$
of the two parameters associated with
the vortex scattering. Note that for $b_{\kappa}=b_{\sigma}\rightarrow0$,
we recover  the previous result of $L/L_{00}=1/2$.
In our previous work\cite{kim2} 
we were able to determine $b_{\kappa}$ from the saturated value
of the thermal conductivity as a function of a magnetic field, and so
Eq.~(\ref{L_b}) would give $b_{\sigma}$ directly from the Lorenz number.

\begin{figure}[tp]
\begin{center}
\includegraphics[height=2.4in]{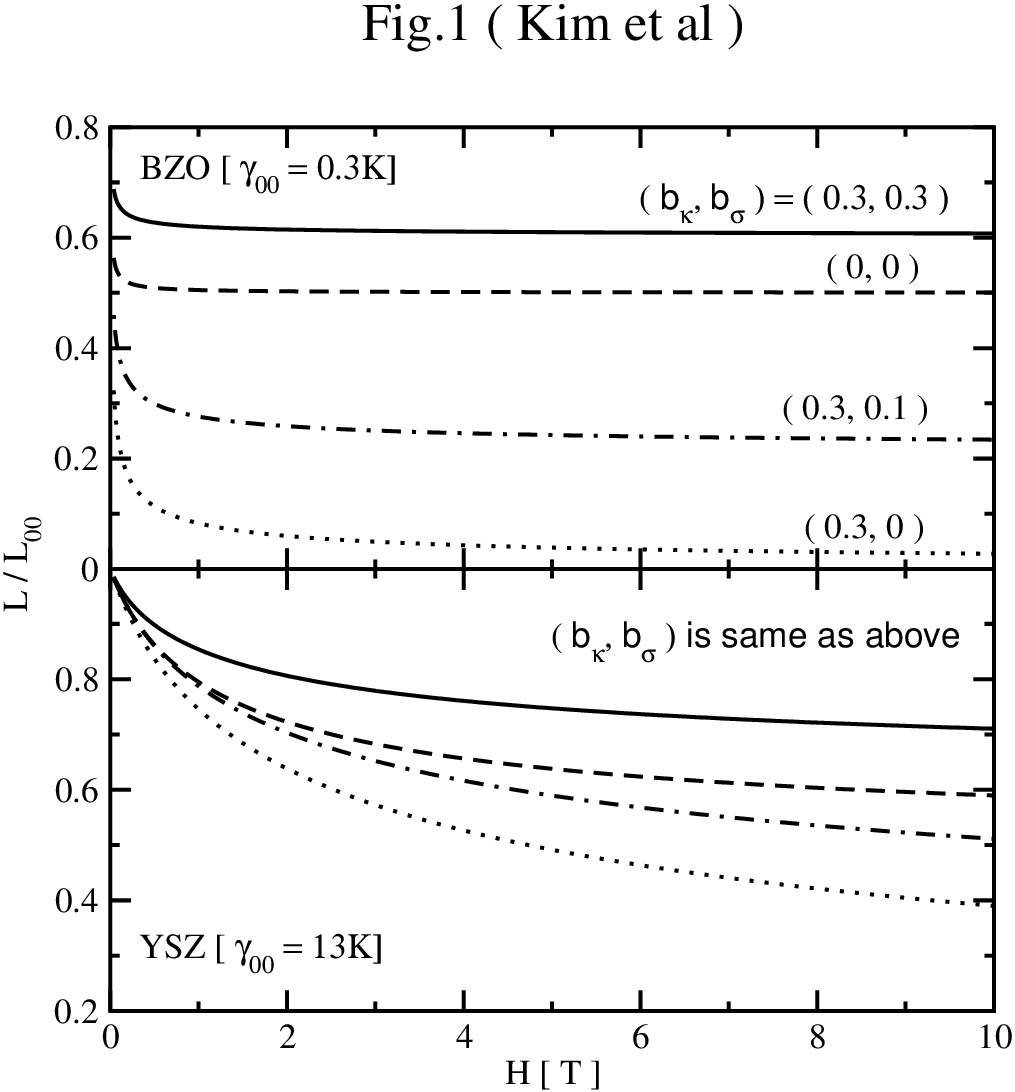}
\caption{
The Lorenz number $L$ normalized to its universal value $L_{00}\simeq\pi^{2}/(3e^{2})$
as a function of a magnetic field $H$ applied perpendicular to the CuO$_{2}$ plane.
The top frame is for a pure sample (BZO) with $\gamma_{00}=0.3K$ while the bottom
frame is for a dirtier sample (YSZ) with $\gamma_{00}=13K$.
(See Refs.\cite{hill2,kim2} for details.)
Various choices of the
vortex scattering parameters $b_{\kappa}$ and $b_{\sigma}$ are considered:
$(b_{\kappa},b_{\sigma})=(0.3,0.3)$ for the solid curves in both frames,
$(0,0)$, $(0.3,0.1)$, and $(0.3,0)$ for the dashed, dot-dashed, and dotted curves,
respectively.
}
\end{center}
\end{figure}

We performed numerical calculations of the Lorenz number 
for various values of $b_{\kappa}$ and $b_{\sigma}$ at $T=0$. In Fig.~1, we plot 
results for $L/L_{00}$ as a function of $H$ in two specific cases.
The top frame is for $\gamma_{00}=0.3K$, a value of the impurity scattering rate deduced 
from consideration of the thermal conductivity
of an ultra pure sample\cite{hill2,kim2}
of YBCO$_{6.99}$ denoted by BZO and the bottom frame is for a dirty sample
denoted by YSZ with $\gamma_{00}=13K$. The values of $(b_{\kappa},b_{\sigma})$
are shown as labels to the curves. For example, $(b_{\kappa},b_{\sigma})=(0.3,0.3)$
for the solid curve. For the dashed, dot-dashed and dotted curve
$(b_{\kappa},b_{\sigma})=(0,0),(0.3,0.1)$, and $(0.3,0)$, respectively.
Note that the analytic result Eq.~(\ref{half_L}) is verified by the dashed
curve with $(0,0)$ in the top frame. In fact, in this case, the value $1/2$ is reached
already around $H=1$ Tesla. For the bottom frame (YSZ sample)
the reduction to $1/2$ is not yet seen even at $10$ Tesla
because the saturation is controlled not by $H$ but by the ratio 
$E_{H}/\gamma_{00}$, which is needed to be much larger than $1$ in order to
realize Eq.~(\ref{half_L}).
Estimates of $E_{H}/\gamma_{00}$ are about $30\sqrt{H}$ (Tesla$^{-1}$) for the top frame
which is much larger than $1$ for $H=1$ Tesla. But the bottom sample is dirtier
by a factor of approximately $40$. Eq.~(\ref{L_b}) is also verified in our numerical
work. Next we turn to the consideration of the effects of a magnetic field on the 
impurity scattering. This takes us beyond the simple 
Matthiessen's rule which we relied on in this section.


\section{magnetic field effect on impurity scattering}

We now return to the impurity self energy given by Eq.~(\ref{Sigma}). 
So far, for simplicity, we have treated the self energy approximately, by
appealing to Matthiessen's rule; we simply added
the Andreev scattering effect $bE_{H}$ to the impurity scattering
$\gamma$ which is itself left unaltered by the magnetic field.
Here $b=b_{\kappa(\sigma)}$ for the thermal (dc) conductivity
as we mentioned.
Of course, in reality the existence
of a vortex lattice, which introduces spatial inhomogeneities, will
modify the impurity term itself in Eq.~(\ref{g_tot}). We need to return
to Eq.~(\ref{Sigma}) for the retarded self energy and include the effects of
the magnetic field on $G_{0}$ itself. In obtaining Eq.~(\ref{Sigma})
an impurity average has been performed in the standard way to restore
translational invariance. After this procedure 
is carried out, the momentum becomes a good
quantum number again and one can use the semiclassical approximation to take into
account the Doppler shift. 
This re-introduces spatial inhomogeneities. A second average over
a random vortex distribution is then taken.
The legitimacy behind such separate averaging
procedures is an assumption that the impurities and vortices are uncorrelated.

To deal with both impurity and vortex averaging in a
manageable
way, what we propose here is to replace $G_{0}(\omega)$ in Eq.~(\ref{Sigma})
with
its vortex averaged counterpart.
Returning to Eq.(\ref{G_0})
valid at low T, let us introduce the Doppler shift and take an average
over, for example, the Gaussian distribution. Then 
in the Born limit, we obtain for the
total scattering rate $\gamma_{tot}(\omega)$ at $\omega=0$ 
using $\Sigma_{i,ret}(E_{H})=-i\gamma(E_{H})$
\bwt
\be
\gamma_{tot}=4\eta_{B}\int{}d\epsilon{\cal P}(\epsilon)\left[
\gamma_{tot}\ln\left(\frac{p_{0}}{\sqrt{\epsilon^{2}+\gamma^{2}_{tot}}}\right)
+\epsilon\arctan\left(\frac{\epsilon}{\gamma_{tot}}\right)\right]+bE_{H}\;,
\label{g_tot2}
\ee
\ewt
where $\eta_{B}=n_{i}V^{2}_{i}/\left(2\pi v_{f}v_{g}\right)$. 
The factor $4$ in Eq.~(\ref{g_tot2})
appears for isotropic impurity scattering. When we consider
the anisotropic scattering, this factor will be changed to reflect
the scattering anisotropy. Eq.~(\ref{g_tot2}) is to be solved self-consistently
for the effective total scattering rate $\gamma_{tot}(E_{H})=\gamma(E_{H})+bE_{H}$.
Results for the impurity scattering $\gamma(E_{H})$ are plotted in Fig.~2, where
$\gamma(E_{H})$ is normalized to the zero frequency and zero
field scattering rate $\gamma(0,0)=\gamma_{00}$.
The magnetic energy in Fig.~2 is also normalized to $\gamma_{00}$. The top frame
applies to the Born limit and the bottom to the unitary limit. Three cases are
considered. The solid curve is for $\eta_{B}=0.04$ and $b=0.1$, the dashed curve
for $\eta_{B}=0.08$ and $b=0.1$. The dotted curve is for $b=0$ and applies to all
values of $\eta_{B}$, as is shown below. 
After some simple algebra we find from Eq.~(\ref{g_tot2})
\bwt
\be
4\eta_{B}\int{}d{\bar\epsilon}{\cal P}({\bar\epsilon})\left[
{\bar\gamma}_{tot}
\ln\left(\frac{1}{\sqrt{{\bar\epsilon}^{2}+{\bar\gamma}^{2}_{tot}}}\right)
+{\bar\epsilon}
\arctan\left(\frac{{\bar\epsilon}}{{\bar\gamma}_{tot}}\right)\right]+
b{\bar E}_{H}=0\;,
\label{g_tot3}
\ee
\ewt
where ${\bar\epsilon}=\epsilon/\gamma_{00}$, ${\bar\gamma}_{tot}=
\gamma_{tot}/\gamma_{00}$, and ${\bar E}_{H}=E_{H}/\gamma_{00}$.
It is clear from Eq.~(\ref{g_tot3}) that for $b=0$, the resulting equation
does not depend on the value $\eta_{B}$. Consequently, the dotted
curve in the top frame of Fig.~2 results regardless of a specific value
of $\eta_{B}$. For any finite $b$, however, the result does depend on $\eta_{B}$.
We note that in all cases in the Born limit, the effect of the magnetic field $H$
is to increase the scattering rate, and that for a 
given value of $E_{H}/\gamma_{00}$ the increase
is largest for the smallest $\eta_{B}$ value. These results confirm the expectation 
based on the following argument.
Let $\Gamma_{N}$ be the impurity scattering rate in the
normal state. For the Born limit, the zero frequency scattering rate
in the superconducting state is much less than $\Gamma_{N}$ while
for the unitary limit it is larger. In fact, for the Born limit
the decrease in $\gamma_{00}$ is exponentially dependent on $\Gamma_{N}$.
As we introduce more vortices by increasing the magnetic field,
more normal regions appear at the expense of superconducting ones and,
hence, we would expect that, on average, the effective scattering rate
would increase towards $\Gamma_{N}$. This is precisely what we
have found. On the other hand,
in the unitary limit $\gamma_{00}$ is very much larger than $\Gamma_{N}$.
A useful and well-known approximate formula\cite{hirschfeld1,hirschfeld2} 
for a relation between these two
quantities in this limit is $\gamma_{00}\simeq 0.63\sqrt{\Gamma\Delta_{0}}$. 
Thus, as
$H$ is increased we expect now that $\gamma(E_{H})$ will decrease because it should
tend towards its normal state value. This expectation is confirmed in the
bottom frame of Fig.~2, where we present numerical results for two
values of $b$; namely, $b=0$ and $b=0.1$, and two values of 
$\eta_{U}=2\pi n_{i}v_{f}v_{g}/\gamma^{2}_{00}$
($12$ and $20$). One can estimate
$\eta_{U}$ using $\Gamma=n_{i}/\pi N(0)$, $\gamma_{00}\simeq 0.63\sqrt{\Gamma\Delta_{0}}$,
and $\Delta_{0}\simeq\pi N(0)v_{f}v_{g}$ in the nodal approximation.
In all cases, as expected $\gamma(E_{H})$ decreases and the decrease is larger
when $b\ne0$. The results in the unitary limit are based on the following equation:
\bwt
\be
\gamma_{tot}=\frac{\eta_{U}}
{4\int{}d\epsilon
{\cal P}(\epsilon)\left[
\gamma_{tot}\ln\left(\frac{p_{0}}{\sqrt{\epsilon^{2}+\gamma^{2}_{tot}}}\right)
+\epsilon\arctan\left(\frac{\epsilon}{\gamma_{tot}}\right)\right]}
+bE_{H}\;.
\ee
\ewt
One can see that, in this case, even for $b=0$, $\gamma(E_{H})$ still
depends on $\eta_{U}$. These expressions are used to obtain results 
at the bare bubble level for
the dc and thermal conductivity as a function of $E_{H}$ (see Figs.~3 and 4
below for $b=0$ and $b=0.3$, respectively).

Even when Andreev scattering from randomly distributed vortices is not considered,
we still see, in Fig.~3, that including the effect of $H$ in $\gamma_{tot}$
increases both $\kappa$ (top frame) and $\sigma$ (middle frame)
over the values (dotted curve) it would have if $\gamma_{00}$ was used 
instead of $\gamma(E_{H})$ (i.e. no effect of $H$ is included in $\gamma$) 
in the unitary limit while the opposite holds in the Born limit.
This is as expected from Fig.~2, where it was shown that $H$ reduces
$\gamma(E_{H})$ in the unitary limit while it increases 
$\gamma(E_{H})$
in the Born limit. Further, in Fig.~4, where a non-zero value of $b=0.3$
is considered, the effects remain as in Fig.~3 but all variations with
$E_{H}$ are much reduced. This means that Andreev scattering off
vortices offsets to a large degree the increased
conductivity due to the increase of quasiparticles
created by the Doppler shift. Returning to the bottom frames of Figs.~3
and 4, we note that in all cases the Lorenz number is reduced below
its universal value $L_{00}$ as the field is increased. In particular
for $b=0$, the decrease is very rapid, and at $E_{H}/\gamma_{00}=2$
the unitary case has dropped by more than $30\%$
while in the Born limit it is by more than $20\%$. Also for this
ratio, including the effect of $H$ on the impurity scattering is not important 
in the unitary limit but is more important in the Born case. This
also holds for $b=0.3$ shown in the bottom frame
of Fig.~4. There is no qualitative difference between the $b=0$ and $b=0.3$
cases but the variations are not as rapid when Andreev scattering
is included in the calculations.

\begin{figure}[tp]
\begin{center}
\includegraphics[height=2.6in]{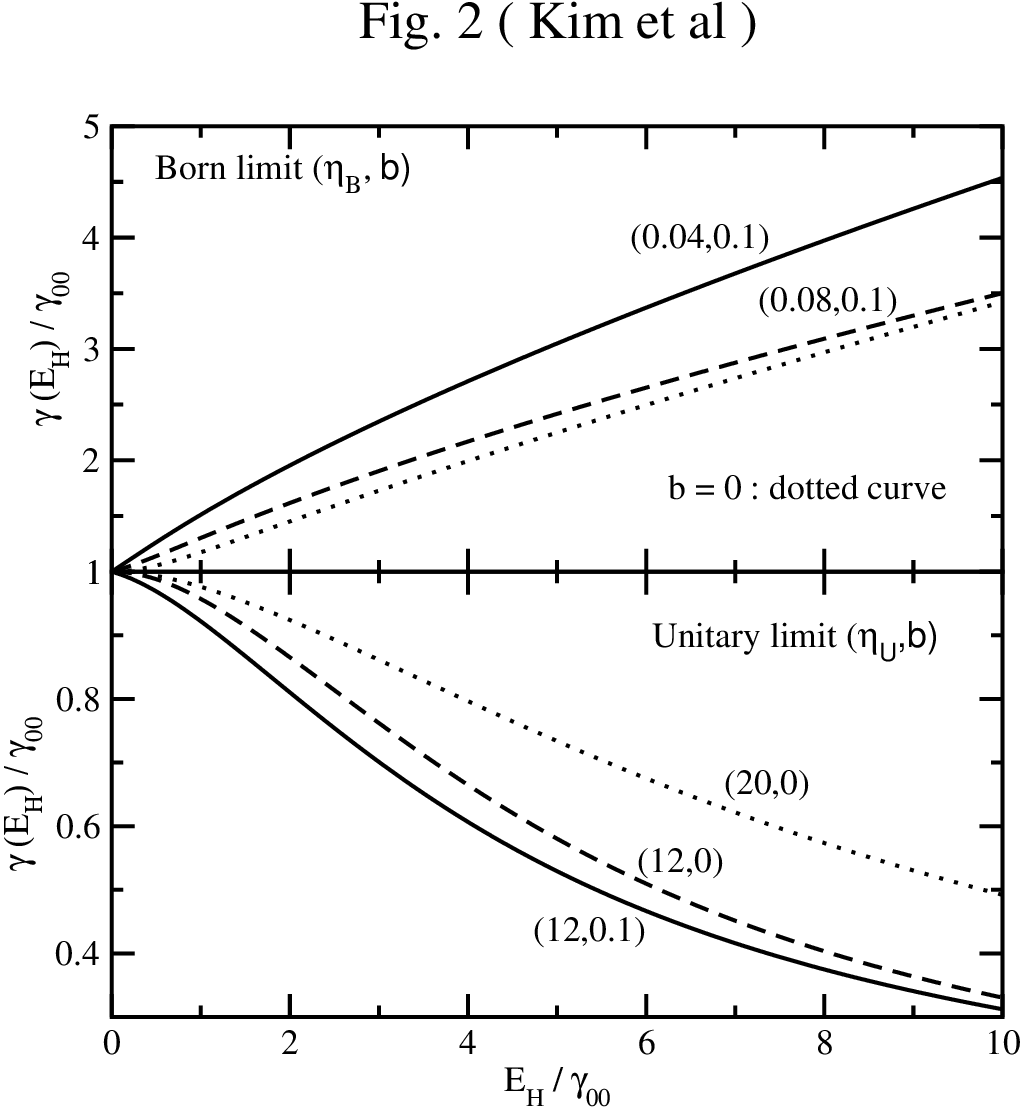}
\caption{
The magnetic field dependence of the impurity scattering rate $\gamma(E_{H})$
as a function of the magnetic energy $E_{H}\propto \sqrt{H}$. The normalization factor
for $\gamma$ and $E_{H}$ is $\gamma_{00}$. The top frame is for the Born limit
with $\eta_{B}=n_{i}V^{2}_{i}/(2\pi v_{f}v_{g})$. The values of $\eta_{B}$ are 
$0.04$ and $0.08$ as labelled. The vortex scattering parameter is chosen to be $0$ or $0.1$.
Note that
for $b=0$, the value of $\eta_{B}$ is irrelevant for $\gamma(E_{H})$ in the Born limit.
The bottom frame is for the unitary limit with $\eta_{U}=2\pi n_{i}v_{f}v_{g}/\gamma^{2}_{00}$.
The values of $\eta_{U}$ are chosen to be $12$ or $20$, and $b=0$ or $0.1$.
}
\end{center}
\end{figure}
\begin{figure}[tp]
\begin{center}
\includegraphics[height=2.6in]{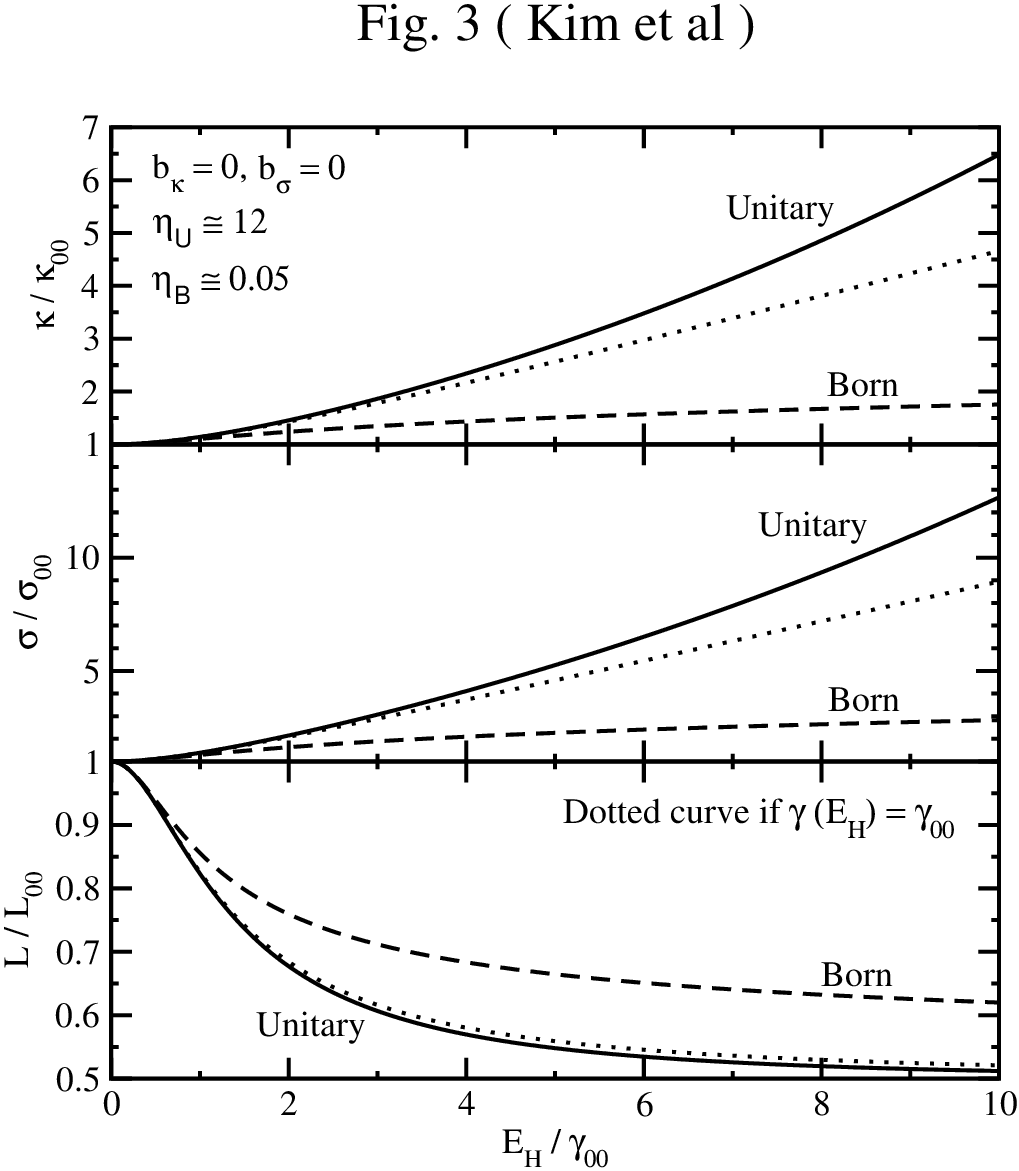}
\caption{
The thermal (top), dc (middle) conductivity and
Lorenz number (bottom frame) as a function of the magnetic
energy $E_{H}$. 
All calculations are based on the bare bubble diagram; however, the magnetic
field dependence of the impurity scattering
is included.
The vortex scattering is neglected $(b_{\kappa}=b_{\sigma}=0)$.
The solid curves in all frames are for the unitary limit $(\eta_{U}\simeq12)$ while
the dashed curves are for the Born limit $(\eta_{B}=0.05)$. The dotted curves are for the case
when the magnetic field dependence of the impurity scattering is neglected; namely,
$\gamma_{00}$ is used instead of $\gamma(E_{H})$ in the self energy.
}
\end{center}
\end{figure}
\begin{figure}[tp]
\begin{center}
\includegraphics[height=2.6in]{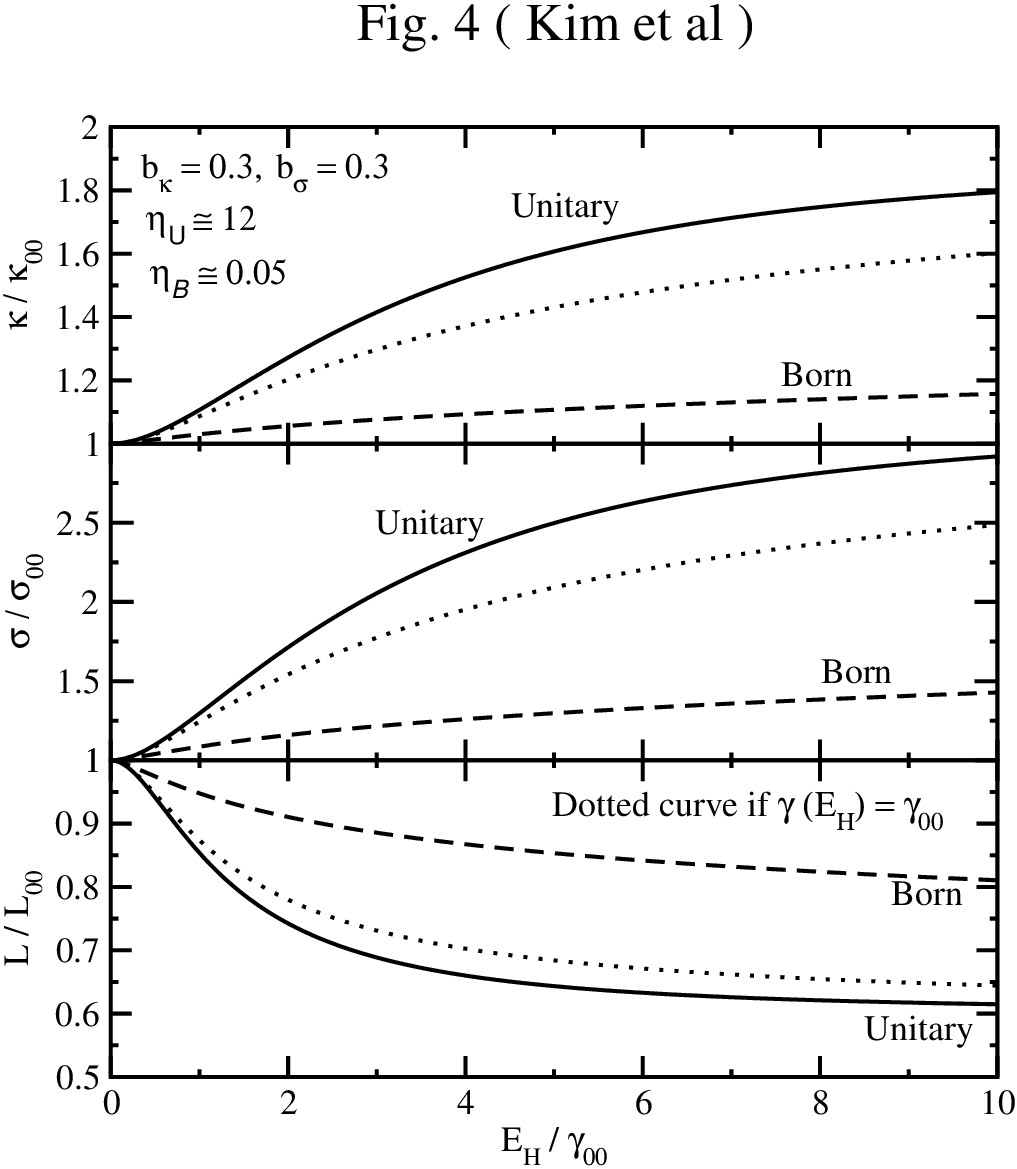}
\caption{
The thermal (top), dc (middle) conductivity 
and the Lorenz number (bottom frame) as a function of the magnetic
energy at the bare bubble level including
the field dependence of the impurity scattering.
In this figure, the Andreev scattering is considered $(b_{\kappa}=b_{\sigma}=0.3)$.
Other parameters are the same as in Fig.~3.
}
\end{center}
\end{figure}

\section{vertex corrections to conductivity}

We now consider the effect of vertex corrections to the dc and thermal 
conductivity of a $d$-wave superconductor in the vortex state.
This problem was treated in detail by DL in the case 
when the impurity scattering is anisotropic but no magnetic field is applied
to the sample. 
The necessary mathematics is rather lengthy so that here
we will deal only with the essential modifications to the formalism of Ref.\cite{durst}
that are required when a magnetic field is applied. 
Following Durst and Lee, we also introduce impurity scattering potentials
to the same node $(V_{1})$, the adjacent node $(V_{2})$, and
the opposite node $(V_{3})$. Thus the anisotropy can be
parametrized by $R_{2}=V_{2}/V_{1}$ and $R_{3}=V_{3}/V_{1}$. If
$R_{2}=R_{3}=1$, the impurity scattering is isotropic.
Our aim here will not be
to give the most general formulation but rather consider a simplified case which can
nevertheless provide a preliminary understanding of the effects of the magnetic field on the
vertex corrections to the conductivity. The most important modification in the formalism
is to replace, for all relevant calculations, the Green's function by its average
over the vortex distribution. This replacement may limit the generality
of our results but allows us to get a concrete numerical evaluation of the
resulting vertex corrections. A central function in the calculations is $F(z)$
as in Ref.\cite{durst} where
\be
F(z)=-4\pi v_{f}v_{g}
\int^{p_{0}}_{0}\frac{d^{2}p}{(2\pi)^{2}v_{f}v_{g}}{\hat G}(p,z)
\ee
with $z=i\omega-\Sigma_{i}-\Sigma_{vor}$. Note that
$\Sigma_{vor}$ is the self energy due to vortices so that it includes both
effects of the Doppler shift and Andreev scattering.

Let us consider the vertex corrections to the dc conductivity.
In terms of $F(z)$ the dc conductivity can be written as 
\bwt
\be
\frac{\sigma_{VC}}{\sigma_{00}}=
\frac{1}{2}
\int{}d\epsilon {\cal P}(\epsilon)~
\mbox{Re}\left[\frac{I^{(0)}_{2}(\epsilon)}{1-A^{(0)}_{2}}
-\frac{I^{(0)}_{1}(\epsilon)}{1-A^{(0)}_{1}}\right]\;,
\label{sigma_ver}
\ee
\ewt
which replaces Eq.~(3.13) of DL.
In the above Eq.~(\ref{sigma_ver}),
\be
I^{(0)}_{1}(\epsilon)=\frac{\partial F(z)}{\partial z}\Bigg|_{z=z_{0}}
\ee
and
\be
I^{(0)}_{2}(\epsilon)=\frac{F(z)-F(z^{*})}{z-z^{*}}\Bigg|_{z=z_{0}}
\ee
with $z_{0}=-\epsilon+i\gamma_{t}(0)$ and
\be
F(z)=2\int^{p_{0}}_{0}pdp\frac{z}{p^{2}-z^{2}}\simeq2z\ln\frac{ip_{0}}{z}\;.
\ee
The parameters $A^{(0)}_{1}$ and $A^{(0)}_{2}$ are due to the scattering
anisotropy. They will be specified later but for isotropic scattering
they vanish. In this instance, we obtain
\be
\frac{\sigma}{\sigma_{00}}=
\frac{1}{2}
\int{}d\epsilon {\cal P}(\epsilon)~
\mbox{Re}\left[I^{(0)}_{2}(\epsilon)-I^{(0)}_{1}(\epsilon)\right]\;,
\ee
which needs to reduce to the expression we obtained in the previous
section where the vertex corrections were neglected.
To see if we recover the bare bubble results we need to calculate
$I^{(0)}_{1}$ and $I^{(0)}_{2}$;
\be
I^{(0)}_{1}(\epsilon)=
2\ln\frac{ip_{0}}{z_{0}}-2
\ee
and
\be
I^{(0)}_{2}(\epsilon)
=\frac{2}{\gamma_{t}(0)}\mbox{Im}\left\{z_{0}
\ln\frac{ip_{0}}{z_{0}}\right\}\;.
\ee
Now we obtain
\be
I^{(0)}_{2}(\epsilon)-I^{(0)}_{1}(\epsilon)
=2-2\frac{\epsilon}{\gamma_{t}(0)}
\mbox{arg}\left(\frac{ip_{0}}{z_{0}}\right)
-2i~\mbox{arg}\left(\frac{ip_{0}}{z_{0}}\right)
\ee
where
\be
\mbox{arg}\left(\frac{ip_{0}}{z_{0}}\right)=
-\arctan\left(\frac{\epsilon}{\gamma_{t}(0)}\right)\;.
\ee
Thus the dc conductivity becomes
\be
\frac{\sigma}{\sigma_{00}}=
\int{}d\epsilon {\cal P}(\epsilon)
\left[1+\frac{\epsilon}{\gamma_{t}(0)}
\arctan\left(\frac{\epsilon}{\gamma_{t}(0)}\right)\right]\;,
\ee
which is what we had before at the bare bubble level
(Eqs.~(\ref{sigma_0H}) and (\ref{s00})).

Now we consider the vertex corrections to the dc conductivity
in the Born limit. In this case, following DL we obtain after
the appropriate generalization
$A^{(0)}_{1}=\alpha^{(0)}_{1}\langle
I^{(0)}_{1}\rangle$ and
$A^{(0)}_{2}=\alpha^{(0)}_{2} \langle I^{(0)}_{2}\rangle$
with $\alpha^{(0)}_{1}=a^{(0)}_{2}$, where $\langle I^{(0)}_{i}
\rangle=\int d\epsilon {\cal P}(\epsilon) I^{(0)}_{i}(\epsilon)$
$(i=1,2)$ and
\be
\alpha^{(0)}_{1}=\frac{n_{i}V^{2}_{1}}{4\pi v_{f}v_{g}}
(1-R^2_{3})\;.
\ee
Note in this case,
\be
\ln\frac{p_{0}}{\gamma(0)}=\frac{1}
{\eta_{B}\left(1+2R^{2}_{2}+R^{2}_{3}\right)}\;,
\ee
where $\left(1+2R^{2}_{2}+R^{2}_{3}\right)\rightarrow4$ for the isotropic scattering
$(R_{2}=R_{3}=1)$ as we mentioned for Eq.~(\ref{g_tot2}) in the previous section.
The expression for the dc conductivity, including the vertex corrections
$\sigma_{VC}$, can be simplified and becomes
\be
\frac{\sigma_{VC}}{\sigma_{00}}=
\frac{1}{2}
\left[\frac{\langle I^{(0)}_{2}\rangle-
\langle I^{(0)}_{1}\rangle}
{\left(1-\alpha^{(0)}_{1}\langle I^{(0)}_{1}\rangle\right)
\left(1-\alpha^{(0)}_{1}\langle I^{(0)}_{2}\rangle\right)}\right]
\ee
Since 
$\frac{\sigma}{\sigma_{00}}=\frac{1}{2}
\left[\langle I^{(0)}_{2}\rangle\
-\langle I^{(0)}_{1}\rangle\right]$, 
we finally obtain
\be
\frac{\sigma_{VC}}{\sigma}=
\frac{1}{
\left(1-\alpha^{(0)}_{1}\langle I^{(0)}_{1}\rangle\right)
\left(1-\alpha^{(0)}_{1}\langle I^{(0)}_{2}\rangle\right)}\;.
\ee 

For the numerical calculations we choose $R_{2}=0.9$ and $R_{3}=0.8$.
Since it is assumed that $V_{1}$ is small, we retain only terms of order
$V^{2}_{1}$. In Fig.~5, we show numerical results
with $\eta_{B}=n_{i}V^{2}_{1}/2\pi v_{f}v_{g}=0.05$.
In a rigorous sense $\eta_{B}\rightarrow0$ since $V_{1}\rightarrow0$
in the Born limit; however, this value is unrealistic for an actual sample.
In the top frame of Fig.~5, we plot $\sigma_{VC}/\sigma_{00}$ as a function of
$E_{H}/\gamma_{00}$ for $b_{\sigma}=0$ (solid curve) and 
$b_{\sigma}=0.1$ (dashed curve). 
At small values of $E_{H}$,
the Andreev scattering does not have a large effect on $\sigma_{VC}/\sigma_{00}$.
For large values of $E_{H}$, however, the difference can be much larger.
The conductivity also rises significantly above its universal value.
This rise is not to be assigned to the vertex corrections. In the bottom frame we plot
$\sigma_{VC}/\sigma$, where $\sigma$ is the dc conductivity at the bare bubble level.
We see now that 
in both cases, including $(b_{\sigma}\ne0)$ and ignoring $(b_{\sigma}=0)$
Andreev scattering, the application of a magnetic field decreases
the effect of the vertex corrections although the decrease is not large
in magnitude. Note that for $H=0$ one can get analytically
$\sigma_{VC}/\sigma=\left(1+2R^{2}_{2}+R^{2}_{3}\right)^{2}/
\left(2R^{2}_{2}+2R^{2}_{3}\right)^{2}$ as in Ref.\cite{durst}
For $R_{2}=0.9$ and $R_{3}=0.8$, $\sigma_{VC}/\sigma\simeq1.26$ for
$H=0$. In our numerical calculation, $\sigma_{VC}/\sigma\simeq1.24$ 
at $H=0$ is a little
less than $1.26$ because, as we mentioned, $\eta_{B}$ is finite.
However, when we reduce $\eta_{B}$ to values smaller than $0.05$, our numerical
results approach this analytic value; for example, for $\eta_{B}=0.01$
we obtain $\sigma_{VC}/\sigma\simeq1.26$.

A more interesting case is the unitary scattering limit, where
the vertex corrections are much larger at zero field. The parameters 
$A^{(0)}_{1}$ and $A^{(0)}_{2}$ are now,
$A^{(0)}_{1}=\langle\beta^{(0)}_{1}\rangle\langle I^{(0)}_{1}\rangle$ and
$A^{(0)}_{2}=\langle\beta^{(0)}_{2}\rangle\langle I^{(0)}_{2}\rangle$ with
$\langle\beta^{(0)}_{2}\rangle
=|\langle\beta^{(0)}_{1}\rangle|$, where $\langle\beta^{(0)}_{1}\rangle
=\int d\epsilon{\cal P}(\epsilon)\beta^{(0)}_{1}(\epsilon)$.
For $R_{2}=0.9$ and $R_{3}=0.8$, we obtain
\be
\ln\frac{p_{0}}{\gamma(0)}=\frac{3n_{i}}{4}
\frac{2\pi v_{f}v_{g}}{\gamma(0)^{2}}
\ee
and
\be
\beta^{(0)}_{1}=\frac{n_{i}}{4}\frac{2\pi v_{f}v_{g}}
{z^{2}_{0}\ln^{2}(ip_{0}/z_{0})}
\ee
where $z_{0}=-\epsilon+i\gamma_{t}(0)$ as before.
Our numerical results are given in Fig.~6. The format is the same as
Fig.~5. The impurity parameter $\eta_{U}=10$ and $b_{\sigma}=0,0.1$.
Referring to the bottom frame we see that the application of a magnetic field
reduces the vertex corrections rapidly as the magnetic field increases up to
a few times $\gamma_{00}$.

For completeness we have also considered
the vertex corrections to the thermal conductivity of a $d$-wave
superconductor in the vortex state. It has been shown that
the vertex corrections are negligible 
at zero field.\cite{durst} In the presence of a magnetic field
we obtain
\bwt
\bea
\frac{\kappa_{VC}}{\kappa_{00}}&=&\frac{1}{2}
\left[1+\left({v_{g}}/{v_{f}}\right)^{2}\right]^{-1}
\int{}d\epsilon{\cal P}(\epsilon)
\mbox{Re}
\left[\frac{I^{(3)}_{2}(\epsilon)}{1-A^{(3)}_{2}}-
\frac{I^{(3)}_{1}(\epsilon)}{1-A^{(3)}_{1}}\right]
\nonumber\\
&+&
\frac{1}{2}
\left[1+\left({v_{f}}/{v_{g}}\right)^{2}\right]^{-1}
\int{}d\epsilon{\cal P}(\epsilon)
\mbox{Re}
\left[\frac{I^{(1)}_{2}(\epsilon)}{1-A^{(1)}_{2}}-
\frac{I^{(1)}_{1}(\epsilon)}{1-A^{(1)}_{1}}\right]\;,
\eea
\ewt
where
\be
I^{(1)}_{2}(\epsilon)=I^{(3)}_{2}(\epsilon)=
\frac{z^{*}F(z)-zF(z^{*})}{(z+z^{*})(z-z^{*})}\Bigg|_{z=z_{0}}
\ee
and
\be
I^{(1)}_{1}(\epsilon)=I^{(3)}_{1}(\epsilon)=
\frac{1}{2}\left[\frac{\partial F(z)}{\partial z}
-\frac{F(z)}{z}\right]_{z=z_{0}}\;.
\ee
Later we will give $A^{(1)}_{1}$ and $A^{(1)}_{2}$
for the Born and unitary limit instead of their
general expressions. For isotropic scattering,
$A^{(1)}_{1}$ and $A^{(1)}_{2}$ vanish again and the thermal
conductivity becomes
\be
\frac{\kappa}{\kappa_{00}}=
\frac{1}{2}\int{}d\epsilon{\cal P}(\epsilon)\mbox{Re}
\left[I^{(1)}_{2}(\epsilon)-I^{(1)}_{1}(\epsilon)\right]\;.
\ee
Since $I^{(1)}_{1}(\epsilon)=-1$ and
\be
I^{(1)}_{2}(\epsilon)=
\left[\frac{\epsilon}{\gamma_{t}(0)}+\frac{\gamma_{t}(0)}{\epsilon}
\right]\arctan\left(\frac{\epsilon}{\gamma_{t}(0)}\right)\;,
\ee
we reproduce the bare bubble result for the thermal conductivity.

In the Born limit, we obtain
$A^{(3)}_{2}=\alpha^{(0)}_{1}\langle I^{(3)}_{2}\rangle$,
$A^{(3)}_{1}=\alpha^{(0)}_{1}\langle I^{(3)}_{1}\rangle$,
$A^{(1)}_{2}=-A^{(3)}_{2}$, and
$A^{(1)}_{1}=-A^{(3)}_{1}$.
Further simple analysis can show that
$1\pm\alpha^{(0)}_{1}\langle I^{(3)}_{2}\rangle\approx1$ and
$1\pm\alpha^{(0)}_{1}\langle I^{(3)}_{1}\rangle\approx1$ since
$\alpha^{(0)}_{1}\sim1/\ln[p_{0}/\gamma(0)]$ and $\epsilon \ll p_{0}$ because
of ${\cal P}(\epsilon)$. In the unitary limit, we obtain
$A^{(3)}_{2}=A^{(1)}_{2}=\langle\beta^{(0)}_{2}\rangle\langle I^{(3)}_{2}\rangle$ and
$A^{(3)}_{1}=A^{(1)}_{1}=\langle\beta^{(0)}_{1}\rangle\langle I^{(3)}_{1}\rangle$. Since
$\langle\beta^{(0)}_{1}\rangle\rightarrow0$ as $p_{0}\rightarrow\infty$,
$1-\langle\beta^{(0)}_{1}\rangle\langle I^{(3)}_{1}\rangle\approx1$ and
$1-\langle\beta^{(0)}_{2}\rangle\langle I^{(3)}_{2}\rangle\approx1$.
We conclude therefore that the vertex corrections to the thermal
conductivity remain negligible even in the
presence of a magnetic field; $\kappa_{VC}\simeq\kappa$. 
In Fig. 7 we plot the Lorenz number,
$L_{\rm VC}/L_{00}$, including vertex corrections, vs. the
normalized magnetic energy. We have used the results for the thermal
conductivity without vertex corrections since they are small, as explained
above. While vertex corrections become less important at high fields,
it is clear form these results that a violation from the WF law remains
at all fields.

\begin{figure}[tp]
\begin{center}
\includegraphics[height=2.6in]{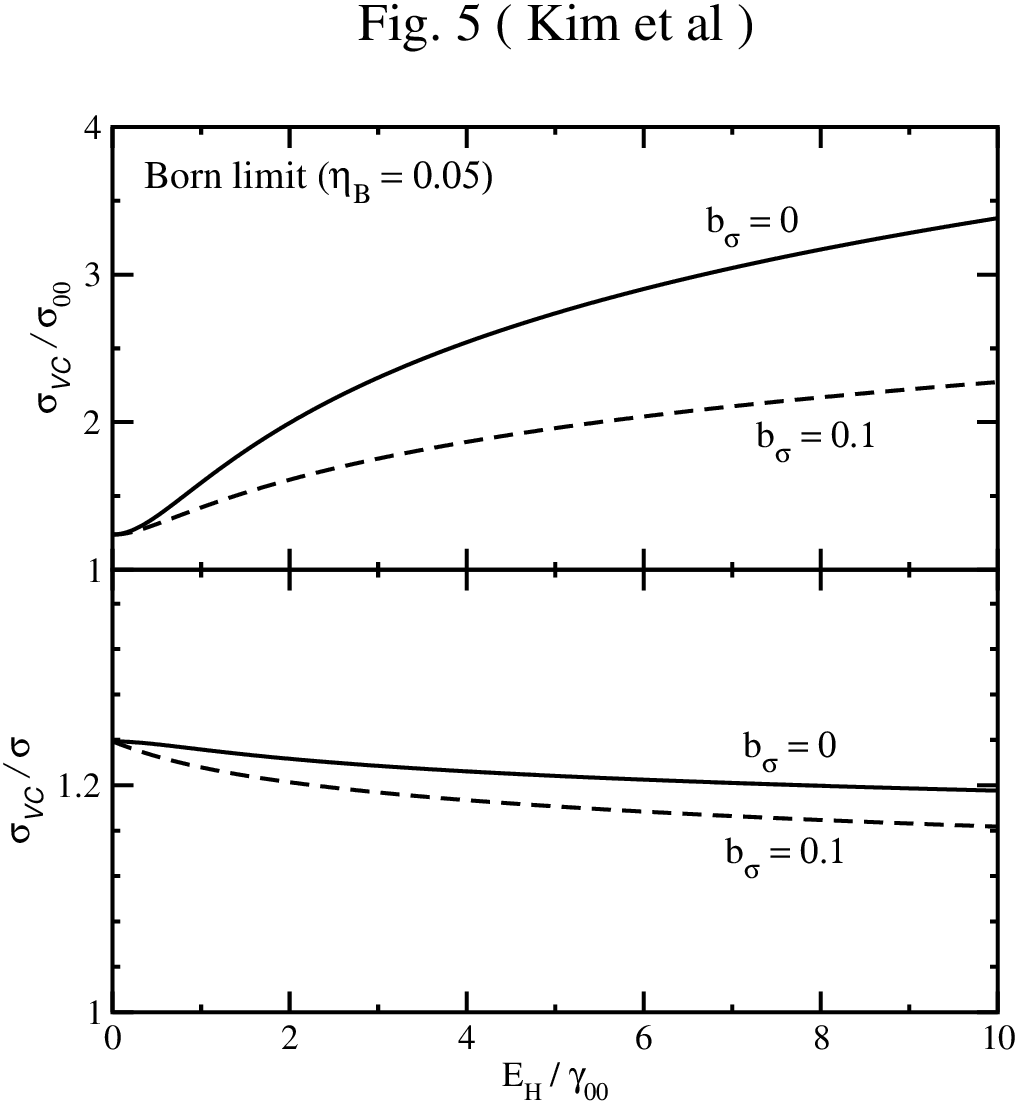}
\caption{
The dc conductivity $\sigma_{VC}$ as a function of $E_{H}/\gamma_{00}$ including 
the vertex corrections 
in the Born limit $(\eta_{B}=0.05)$. The top frame shows $\sigma_{VC}/\sigma_{00}$
while 
the bottom frame
gives $\sigma_{VC}/\sigma$ {\it i.e.} the ratio of its corrected value $(\sigma_{VC})$ to
the dc
conductivity $(\sigma)$ without vertex corrections.
The solid (dashed) curves in both frames are for $b_{\sigma}=0$ $(b_{\sigma}=0.1)$. 
}
\end{center}
\end{figure}
\begin{figure}[tp]
\begin{center}
\includegraphics[height=2.6in]{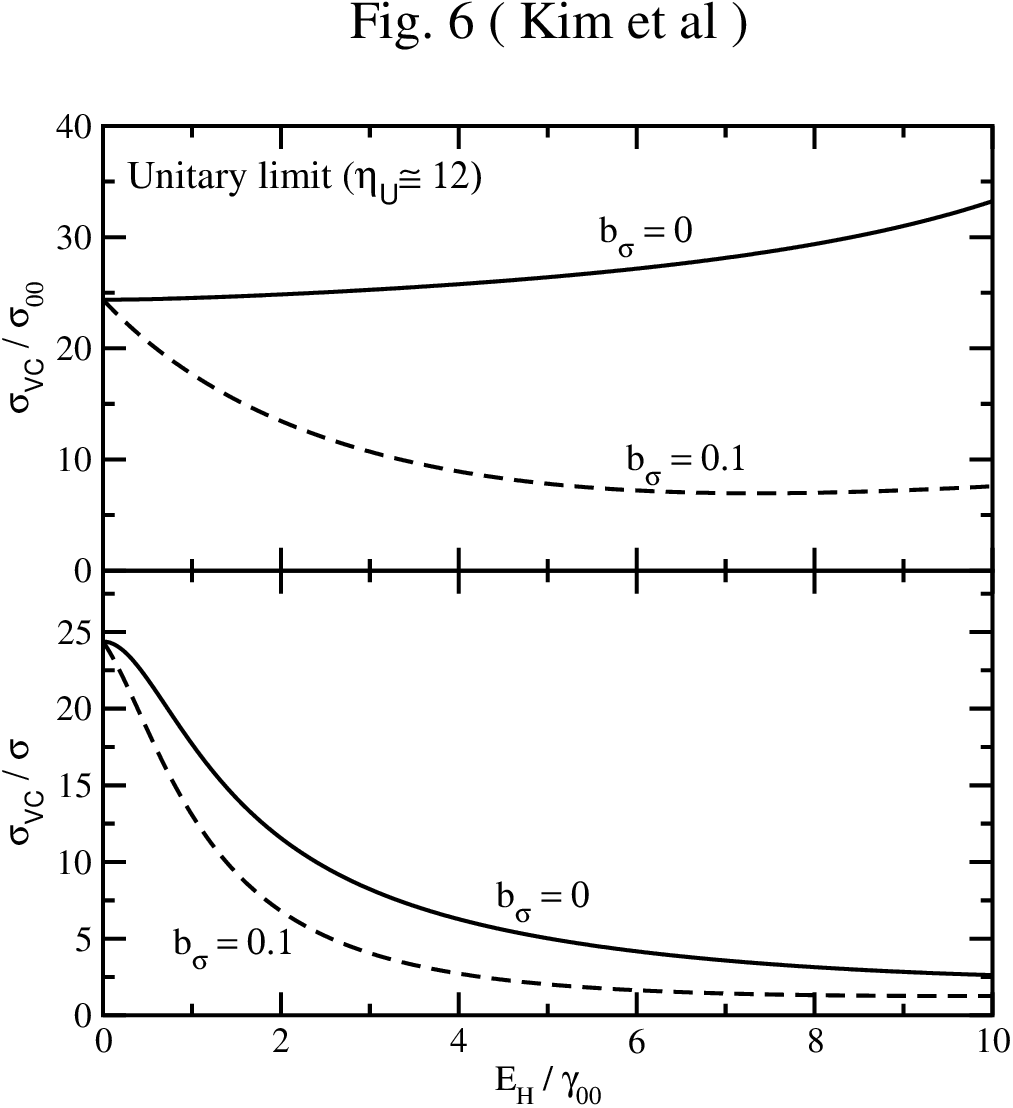}
\caption{
The dc conductivity $\sigma_{VC}$ as a function of $E_{H}/\gamma_{00}$ including
in the unitary limit $(\eta_{U}\simeq12)$. The top frame shows $\sigma_{VC}/\sigma_{00}$
while the bottom frame 
gives $\sigma_{VC}/\sigma$ {\it i.e.} the ratio of its corrected value $(\sigma_{VC})$ to
the dc
conductivity $(\sigma)$ without vertex corrections. 
The solid (dashed) curves in both frames are for $b_{\sigma}=0$ $(b_{\sigma}=0.1)$.
}
\end{center}

\begin{center}
\includegraphics[height=2.6in]{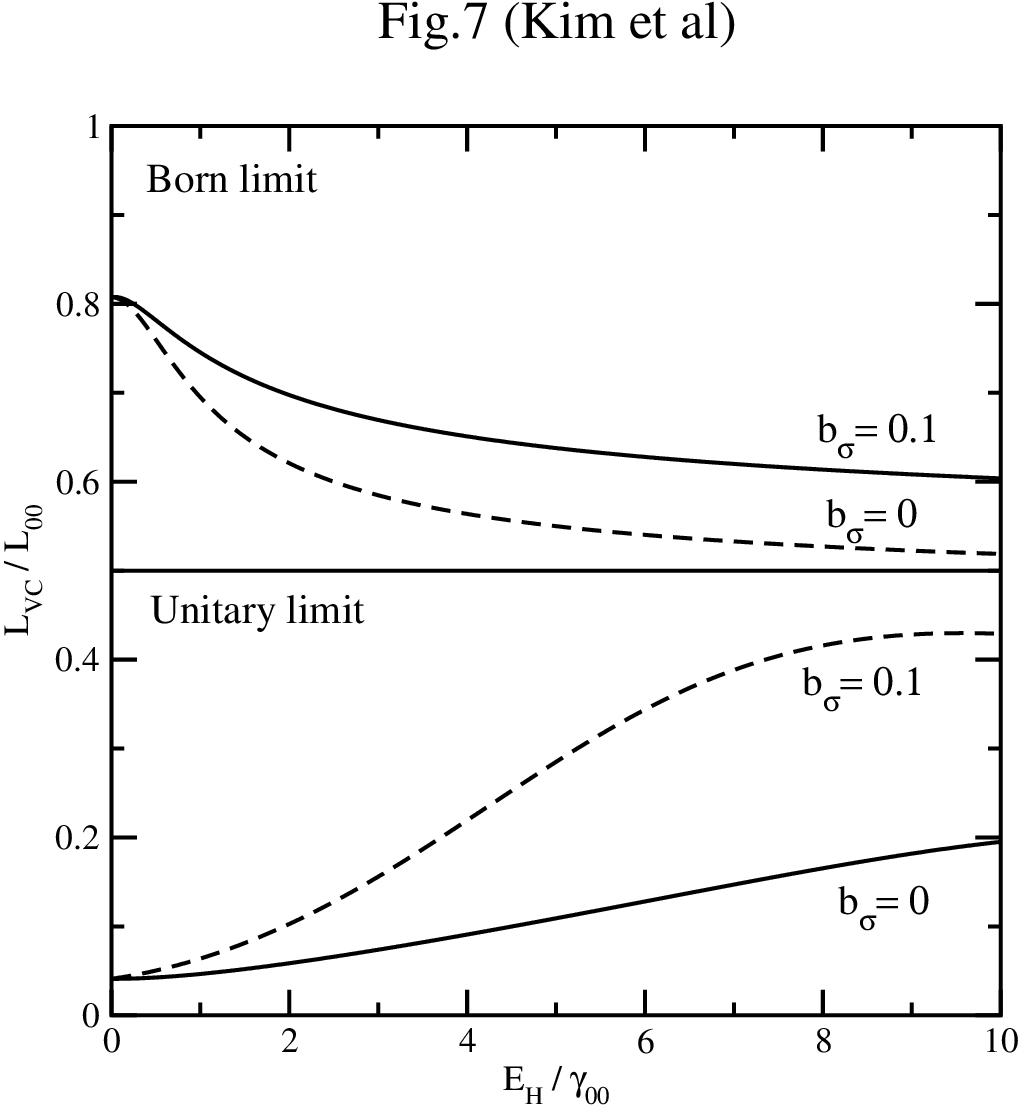}
\caption{
The WF ratio, with vertex corrections as a function of $E_{H}/\gamma_{00}$, for the
cases considered in Fig. 5 and Fig. 6
with $b_{\kappa}=b_{\sigma}$.
The top (bottom) frame is for the Born (unitary)
limit. 
In both cases a violation remains at all magnetic
fields.
}
\end{center}
\end{figure}

\section{conclusions}

We have found that the Wiedemann-Franz law is violated in the mixed state
of a $d$-wave superconductor at zero temperature at the bare bubble
level. Recall that in this approximation there is no
violation in the absence of a magnetic field.\cite{durst}
The deviation
from the universal value of the Lorenz number $L_{00}$ increases with
increasing magnetic field. When Andreev scattering from
vortices is neglected, the high field saturated value of the Lorenz number
can be as low as $L_{00}/2$. With Andreev scattering, it can be even smaller.
We have also found that application of a magnetic field affects
the effective impurity scattering rate $\gamma$
at zero temperature. In the Born limit, $\gamma$ increases while in the 
unitary limit, it decreases as is expected. Physically, in both cases,
$\gamma$ should tend toward its normal state value as the field approaches
$H_{c2}$. In the superconducting state with zero field, the relevant
scattering rate (as is well-known) is much smaller than its normal
state value in the Born limit while the opposite holds in the unitary
case.

We have also considered the vertex corrections arising from
anisotropy in the impurity scattering. It has been shown that in the absence of
a magnetic field the vertex corrections can be large 
for  the zero frequency electrical conductivity at zero temperature
while they are small for the thermal conductivity. We found that application
of the magnetic field reduces the contribution of the vertex corrections
to the electrical conductivity and they become unimportant.
For the thermal conductivity, the corrections remain negligible as in the
absence of a magnetic field.

\begin{acknowledgments}
Work supported in part by the Natural Sciences and Engineering
Research Council of Canada (NSERC), by ICORE (Alberta), and
by the Canadian Institute for Advanced Research (CIAR).
\end{acknowledgments}

\bibliographystyle{prb}

\end{document}